\documentclass[runningheads,a4paper]{llncs}

\usepackage{amssymb}
\usepackage{graphicx}
\usepackage{amsmath}
\usepackage{pgfplots}
\usepackage{epic}
\usepackage{url}
\long\def\remove#1{}
\begin{document}


\title{Geometric graphs on convex point sets}

\titlerunning{Geometric graphs on convex point sets}

\author{Abhijeet Khopkar, Sathish Govindarajan}
\authorrunning{Abhijeet Khopkar and Sathish Govindarajan}
\institute{Department of Computer Science and Automation\\
Indian Institute of Science, Bangalore}
\toctitle{Lecture Notes in Computer Science}
\tocauthor{Authors' Instructions}
\maketitle

\begin{abstract}
In this note, we introduce a family of bipartite graphs called path restricted ordered bipartite graphs and present it as an abstract generalization of some well known geometric graphs like unit distance graphs on convex point sets.
In the framework of convex point sets, we also focus on a generalized version of Gabriel graphs known as locally Gabriel graphs or $LGGs$. $LGGs$ can also be seen as the generalization
of unit distance graphs. The path restricted ordered bipartite graph is also a generalization of $LGGs$. We study some structural properties of the path restricted ordered bipartite graphs and
also show that such graphs have the maximum edge complexity of $\theta(n \log n)$. It gives an alternate proof to the well known result that $UDGs$ and $LGGs$ on convex points have $O(n \log n)$ edges.
\end{abstract}

\section{Introduction}
Tur{\'a}n type problems have a rich history in graph theory. Tur{\'a}n's classical problem is to find the maximum number of edges $ex(n,H)$  a graph (on $n$ vertices) can have without containing
a subgraph isomorphic to $H$ (refer to Tur{\'a}n's theorem ~\cite{turan}). A simple example is that a graph not containing any cycle (acyclic graph) has linear number of edges.
These type of problems have been extensively studied for the geometric graphs~\cite{suk,toth,tth,crss,pv}. 
Various geometric graphs have been studied for special point sets like points on a uniform grid and the convex point sets. 
Tur{\'a}n type problems have also been studied on the geometric graphs when all the points are in convex position~\cite{Capoy,ram,perles,sas,fox}.
The vertices in convex position also provide a cyclic ordering on the vertices. Thus, an obvious technique to explore these problems is by extracting abstract combinatorial structures from geometric conditions
and the order on the vertices.
Similar problems have been addressed by an alternate approach of counting the maximum number of 1s in a 0-1 matrix 
where some sub matrices are forbidden. A 0-1 matrix can represent the adjacency matrix of a bipartite graph with an ordering on the vertices in both the partitions. The maximum number of 1s corresponds to the maximum number of
edges the graph can have. The problem of counting 1s in a 0-1 matrix for various forbidden sub matrices is explored extensively~\cite{furedi,H92,pet10,Keszegh09}.

For many geometric graphs, the edge complexity is studied by characterizing a forbidden subgraph by some geometric restriction, for examples refer to~\cite{furedi,yyy,ps04}.
In this paper, we study {\it Unit distance graphs} and {\it Locally Gabriel graphs} on convex point sets. We characterize some forbidden patterns in these graphs and use them to study the edge complexity.
\subsection{Unit distance graphs}
Unit distance graphs ($UDGs$)\footnote{Not to be confused with the unit disk graphs} are well studied geometric graphs. In these graphs an edge exists between two points if and only if the Euclidean distance between the points is unity. 
\begin{definition}
 A geometric graph $G=(V,E)$ is called unit distance graph provided that for any two vertices $v_1,v_2 \in V$, the edge $(v_1,v_2) \in E$ if and only if the Euclidean
distance between $v_1$ and $v_2$ is exactly unity.
\end{definition}
UDGs have been studied extensively for various properties including their edge complexity.The upper bound and the lower bound for the
number of the maximum edges in the unit distance graphs (on $n$ points in $\mathcal{R}^2$) are $O(n^\frac{4}{3})$~\cite{bb} and $O(n^\frac{1}{c\log\log n})$ (for a suitable constant $c$) respectively~\cite{Erdudg}
Erd\H{o}s showed an upper bound of $O(n^\frac{3}{2})$~\cite{Erdudg}. The bound was first improved to $o(n^\frac{3}{2})$~\cite{szem}, then improved to $n^{1.44\ldots}$~\cite{beck}.
Finally, the best known upper bound of $O(n^\frac{4}{3})$ was obtained by~\cite{bb}. Alternate proofs for the same bound were given by~\cite{Szk,pach}.
Bridging the gap in these bounds has been a long time open problem. Unit distance graphs have also been studied for various special point sets most notably the case when 
all the points lie in convex position. The best known upper bound for the number of edges in a unit distance graph on a convex point set with $n$ points is $O(n \log n)$.
The first proof for this upper bound was given by Zolt{\'a}n F{\"u}redi~\cite{furedi}. The proof is motivated by characterizing a 3 $\times$ 2 sub matrix that is forbidden in a 0-1 matrix.
The sub matrix is motivated by the definition of $UDGs$ and the convexity of the point set.
It was shown that any such $a \times b$ matrix has at most $a+(a+b) \lfloor \log_{2}b \rfloor$ number of 1s. The argument can be easily extended to show that the adjacency matrix of a $UDG$ on a convex point set of size $n$ has $O(n\log n)$ number of 1s that corresponds to the total number of edges.
Peter Bra{\ss} and J{\'a}nos Pach provided an alternate and simple proof using a simple divide and conquer technique~\cite{bp}.
Another proof for the same bound using another forbidden pattern supplemented by a divide and conquer technique was given in~\cite{sas}. The best known lower bound on the number of unit distances in a convex point set is $2n-7$ for $n$ vertices~\cite{han}.
Bridging the gap in the bounds for this special case has also been an open problem. Some interesting questions on the properties of unit distances in a convex point set are studied in~\cite{Erd86,Fb92}.
Unit distance graphs have also been studied for more special types of convex point sets, e.g centrally symmetric convex point set. Unit distance graphs on
centrally symmetric convex point sets have $O(n)$ edges~\cite{csc}.

\subsection{Locally Gabriel Graphs}
Gabriel and Sokal \cite{ggg} defined the Gabriel graph as follows:
\begin{definition} \label{gdef}
A geometric graph $G=(V,E)$ is called a Gabriel graph if the following condition holds:
For any $u,v \in V$, an edge $(u,v) \in E$ if and only if the disk with $\overline{uv}$ as diameter does not contain any other point of $V$.
\end{definition}

Motivated by applications in wireless routing, Kapoor and Li~\cite{yyy} proposed a relaxed version of Gabriel graphs known as $k$-locally Gabriel graphs.
The edge complexity of these structures has been studied in \cite{yyy,ps04}.
In this paper, we focus on 1-locally Gabriel graphs and call them {\em Locally Gabriel Graphs} ($LGG$s).
\begin{definition}\label{lggdef}
A geometric graph $G=(V,E)$ is called a Locally Gabriel Graph if for every $(u,v) \in E$, the disk with $\overline{uv}$ as diameter
does not contain any neighbor of $u$ or $v$ in $G$.
\end{definition}
The above definition implies that two edges $(u,v)$ and $(u,w)$ where $u,v,w \in V$ {\it conflict} with each other if $\angle uwv~\ge~\frac{\pi}{2}$ or $\angle uvw \ge \frac{\pi}{2}$
and cannot co-exist in an $LGG$, i.e. it is not possible to satisfy the condition  $(u,v) \in E$ and $(u,w) \in E$.
Conversely if edges $(u,v)$ and $(u,w)$ co-exist in an $LGG$,
then $\angle uwv~<~\frac{\pi}{2}$ and $\angle uvw < \frac{\pi}{2}$. We call this condition as {\it LGG constraint}.
In this paper, we explore these graphs on convex point sets.
\subsection{Preliminaries and Notations}\label{pn}
A graph is called an ordered graph when the vertex set of the graph has a total order on it. We consider a bipartite graph when the vertex set in each partition has a total order on its vertices.
Formally, an ordered bipartite graph is $G = (U,V,<_U,<_V,E)$. There are two linear ordered sets $(U,<_U)$ and $(V,<_V)$ of the vertices and $E \subseteq U \times V$.
We define a special family of such bipartite graphs where some structures in these graphs are forbidden. We show that the study of these graphs is motivated by their close relationship
with a special family of geometric proximity graphs on convex point sets. A {\it path} in a graph represents a sequence of the edges s.t. two consecutive edges share a vertex.
A path can be represented as a set of edges.

\begin{definition}
 A path $P$ in the ordered bipartite graph  $G=(U,V,E)$ that  visits the vertices in $U$ and $V$ in the order $u_1,u_2,\ldots,u_k$ and $v_1,v_2,\ldots,v_l$ respectively,
is called a forward path if either $u_1 < u_2 \ldots <u_k$ and $v_1 < v_2 \ldots < v_l$ or $u_1 > u_2 \ldots > u_k$ and $v_1 > v_2 \ldots > v_l$.
\end{definition}
An ordered set represented as $<u_1,u_2>$ for $u_1, u_2 \in U$ denotes all the vertices $u_i$ s.t. $u_1 \le u_i \le u_2$. Similarly, an ordered set $<v_1,v_2>$ for $v_1,v_2 \in V$ denotes all the vertices $v_i$ s.t. $v_1 \le v_i \le v_2$.
The {\it range} of a forward path $P$ that passes through the vertices $u_{a}, u_{b}, v_{c}$ and $v_{d}$ is denoted as $\{<u_{a},u_{b}>,<v_{c},v_{d}>\}$,
represents all the vertices (assume that $u_{a} < u_{b}$ and $v_{c} < v_{d}$) $u_i$ and $v_j$ s.t. $u_{a} \le u_i \le u_{b}$ and $v_{c} \le v_j \le v_{d}$.
An edge $(u_{a}, v_j) ($resp. $(v_{c},u_i))$ is called the {\it back edge} to the forward path $P$ if $v_j \in {<v_{c},v_{d}>} ($resp. $u_i \in <u_{a},u_{b}>)$
and $u_i > u_{a_1} ($resp. $v_j > v_{c_1})$ where $u_{a_1} \in U ($resp. $v_{c_1} \in V)$ is a non terminal vertex in $P$, i.e. this vertex has edges incident to two vertices in $P$.
\begin{definition}\label{prp}
 An ordered bipartite graph $G=(U,V,E)$ is said to satisfy the path restricted property if for any forward path $P$ in $G$, there exists no back edge $e \in E$ to $P$.
\end{definition}
A path-restricted ordered bipartite graph ($PRBG$) is an ordered bipartite graph that satisfies the path restricted property.
Note that a $PRBG$ follows the constraint presented by F{\"u}redi~\cite{furedi}, where it was proved that any bipartite graph
following this constraint has $O(n \log n)$ edges. It also implies that a $PBG$ on $n$ vertices has $O(n \log n)$ edges.

To represent these graphs with a Figure, for convenience the vertices are placed from right to left in the increasing order.
\subsection{Our Contributions}
We establish a relationship between UDGs/LGGS on convex point sets and the path restricted ordered bipartite graphs.
The following are the main results presented in this paper.
\begin{itemize}
 \item We prove some structural properties of the path restricted ordered bipartite graphs and their subgraphs.
 \item We give an alternate and simpler proof (compared to the F{\"u}redi's proof~\cite{furedi}) that a path restricted ordered bipartite graph on $n$ points has at most $n \log n + O(n)$ edges.
It also proves that $UDGs/LGGs$ on convex point sets have at most $2n \log n + O(n)$ edges.
 \item We show that if the length of the longest forward path in a $PRBG$ is at most $k$, then this graph has at most $O(k.n)$ edges.
 \item We give a hierarchy between various graph classes. Notably, we show that the class of $UDGs$ on convex point sets is a strict sub class of the class of $LGGs$ on convex point sets.
\end{itemize}
\section{Obtaining $PRBGs$ from $UDGs/LGGs$}\label{sc2}
In this section, we show that a $UDG/LGG$ on convex a point set can be decomposed into two $PRBGs$ by removing at most linear number of edges.
First, we focus on some fundamental properties of the unit distance graphs on a convex point set. Two points $p_i$ and $p_j$ in a convex point set $P$ are called antipodal points
if there exist two parallel lines $\ell_i$ passing through $p_i$ and $\ell_j$ through $p_j$, such that all other points in $P$ are contained between $\ell_i$ and $\ell_j$.
\begin{lemma}\label{ap}
\cite{bp}Let $G_c = (P_c,E)$ be a unit distance graph on convex point set $P_c$. If $p_i \in P_c$ and $p_j \in P_c$ are two antipodal points, then all but at most $2|P_c|$ edges of $G$
cross the line $\overline{p_ip_j}$. 
\end{lemma}

Let $p_1$ and $p_2$ be two antipodal points in the given convex point set $P_c$ as shown in Figure~\ref{ufig1}. Let us divide $P_c$ into two disjoint subsets $U$ and $V$.
$U$ is the set of points above the line $\overline{p_1p_2}$ and $V$ be the set of the points below this line.
Let the vertices in $U$ and $V$ be $u_1,u_2,\ldots u_n$ and $v_1,v_2,\ldots v_m$ respectively (from right to left).
Remove all the edges that do not cross the line $\overline{p_ip_j}$. Let $E'$ be the set of the remaining edges.
Consider the bipartite graph $G = (U,V,E')$. $E'$ is divided into two disjoint sets $E_1$ and $E_2$ by the following rule.
Consider an edge $(u,v_1)$, let $v_0$ and $v_2$ be the adjacent vertices to $v_1$ in $V$ on left and right side respectively as shown in Figure~\ref{ufig2}. By convexity, it can be observed that either $\angle uv_1v_2$ or $\angle uv_1v_0$ is acute.
If $\angle uv_1v_2$ is acute then put the edge $(u,v_1)$ in $E_1$ else if $\angle uv_1v_0$ is acute then put the edge $(u,v_1)$ in $E_2$. If both the angles are acute, then the edge can be put arbitrarily in either $E_1$ or $E_2$.
 In the graph $G_1 = (U,V,E_1)$, the vertices are ordered as $u_1 < u_2 < \ldots u_n$ in $U$ and
 $v_1 < v_2 < \ldots v_m$ in $V$. The ordering is reversed in the graph $G_2 = (U,V, E_2)$.
\begin{remark}
 In $G_1$ and $G_2$, no two edges intersect in a forward path.
\end{remark}
\begin{figure}[ht]
\begin{minipage}[b]{0.5\linewidth}
\centering
\includegraphics[scale=0.5]{udg10}
\caption{Antipodal points in a convex point set}
\label{ufig1}
\end{minipage}
\hspace{0.5cm}
\begin{minipage}[b]{0.4\linewidth}
\centering
\includegraphics[scale=0.55]{udg0}
\caption{Partition of the edges}
\label{ufig2}
\end{minipage}
\end{figure}
Remove the extreme left edge incident to every vertex $v \in V$ from $G_1$, the resultant graph is called $G'_1$. Similarly, by removing
the extreme right edge for every vertex $v \in V$ in $G_2$, the graph $G'_2$ is obtained. Let $G_{UDG}$ denote the class of the ordered bipartite graphs,
consisting of the graphs $G'_1$ and $G'_2$ that are obtained from the unit distance graphs. It can be assumed w.l.o.g. that $|V| \le |U|$.
Thus, a $UDG/LGG$ on convex a point set can be decomposed into two $PRBGs$ by removing at most $3n$ edges.

Consider the Locally Gabriel graphs on a convex point set. Observe that the Lemma~\ref{ap} holds true for Locally Gabriel graphs too.
Therefore, a bipartition can be obtained similarly by dividing a convex point set along two antipodal points. Consider the bipartite graph between the two partitions.
Similar to $G_{UDG}$, a new graph class $G_{LGG}$ can be defined. 
The procedure to obtain a graph in $G_{UDG}$ (from the $UDG$ on a convex point set) can also be applied to an $LGG$ on a convex point set to obtain a graph in $G_{LGG}$.

\begin{figure} 
\begin{center}
 \includegraphics[scale=0.4]{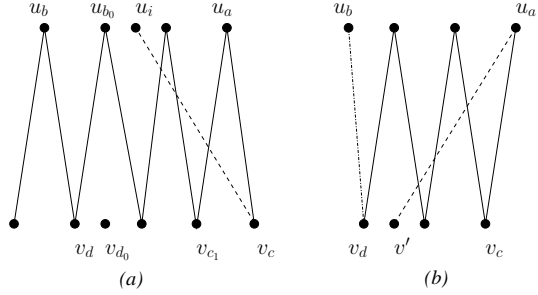}
 \caption{$G_{LGG}$ has path restricted properties}
\label{f3}
\end{center}
\end{figure}

We show that the graphs in $G_{UDG}$ and $G_{LGG}$ are path-restricted ordered bipartite graphs.
\begin{lemma}\label{lggpbg}
 Any graph $G = (U,V,E)$ in $G_{LGG}$ satisfies the path restricted property. Therefore, $G$ is a $PRBG$.
\end{lemma}
\proof
We show that if $P$ is a forward path in $G = (U,V,E)$ with the range $R_P=~\{<u_a,u_b>,<v_c,v_d>\}$, then there does not exist a back edge $(u_i,v_c) \in E$ where $u_i \in <u_a,u_b>$.
The path $P$  and the concerned vertices along with the edges are shown in Figure~\ref{f3}($a$).
Let $v_{d_0} \in V$ be the vertex preceding $v_d$ in $V$. 
Note that $(u_b,v_d)$ is an edge in $P$. Now $\angle u_bv_dv_{d_0} < \frac{\pi}{2}$ (by the definition of $G_{LGG}$).
By convexity, it can be further inferred that $\angle u_bv_dv_c < \frac{\pi}{2}$. 
Let $u_{b_0} \in U$ be the vertex in $P$ with an edge incident to $v_d$ (apart from $u_{b}$) and $v_{c_1} \in V$ be the vertex that immediately succeeds to $v_c$ in $P$. 
By the definition~\ref{lggdef} of $LGGs$, $\angle v_du_bu_{b_0}, \angle u_av_cv_{c_1} < \frac{\pi}{2}$. By convexity, $\angle v_du_bu_a, \angle u_av_cv_d < \frac{\pi}{2}$
Thus, in the quadrilateral $u_av_cv_du_b$, $\angle u_bu_av_c$ must be greater than $\frac{\pi}{2}$. By convexity, $\angle u_iu_av_c > \frac{\pi}{2}$.
Therefore, the edge $(u_i,v_c)$ does not exist in a $G$ for any $u_i \in <u_a,u_b>$.

Recall that the leftmost edge incident to every vertex $v \in V$ is deleted in the graph $G_1=(U,V,E_1)$ to obtain a $G_{LGG}$. Similar arguments lead to the following claim.
If $P$ is a forward path in $G_{LGG} = (U,V,E)$ with the range $R_P = \{<u_a,u_b>,<v_c,v_d>\}$, then there does not exist a back edge $(u_a,v') \in E$ where $v' \in <v_c,v_d>$
(refer to Figure~\ref{f3}($b$)).

Thus, any graph in $G_{LGG}$ satisfies the path restricted property. Therefore, $G_{LGG}$ is a $PRBG$.
\qed
It can be observed that a unit distance graph is also a locally Gabriel graph. Therefore, any graph in the class $G_{UDG}$ also belong to the class $G_{LGG}$. 
\begin{lemma}
 Any graph $G = (U,V,E)$ in $G_{UDG}$ satisfies the path restricted property. Therefore, $G$ is a $PRBG$.
\end{lemma}
\section{Properties of the path restricted ordered bipartite graphs}\label{sc3}
\begin{figure}
 \begin{center}
  \includegraphics[scale=0.35]{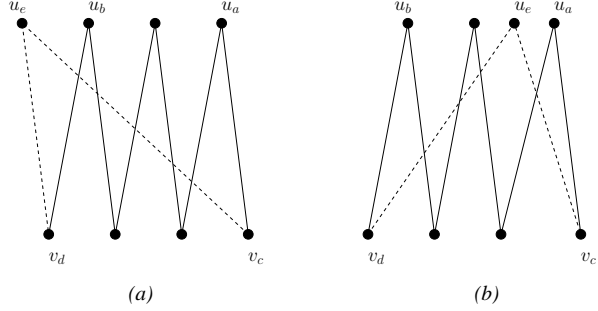}
  \caption{Two edges incident from a vertex to a forward path}
  \label{fcor}
 \end{center}
\end{figure}

In this section, we study some structural properties of $PRBGs$. {\it Path restricted property} results in many interesting structural properties in the ordered bipartite graphs.
\begin{lemma} \label{rem1}
 In a path restricted ordered bipartite graph, two forward paths originating from a vertex in the same direction never meet each other.
\end{lemma}
\begin{figure}
\begin{center}
 \includegraphics[scale=0.4]{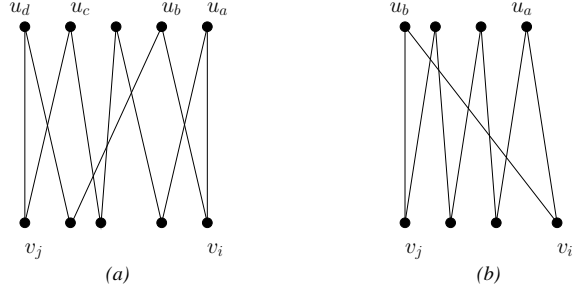}
 \caption{Two forwards paths do not meet in $PRBG$}
\label{ftp}
\end{center}
\end{figure}
\proof
Let us assume that on the contrary, two forward paths $P_1$ and $P_2$ originating from the same vertex meet again as shown in Figure~\ref{ftp}($a$). Let $P_1$ and $P_2$ begin from a vertex $v_i \in V$
and meet at $v_j \in V$. It can be assumed w.l.o.g. that the paths meet in the partition $V$ and $v_i < v_j$. Let $v_i$ has edges incident to $u_a$ (in $P_1$) and $u_b$ (in $P_2$).
It can be assumed w.l.o.g. that $u_a < u_b$ 
Let $v_1$ has edges incident to $u_c$ (in $P_1$) and $u_d$ (in $P_2$).
By the {\it path restricted property}, it is not possible to have $u_b \in <u_a,u_c>$ and $u_b \in <u_a,u_d>$. Thus, it implies that $u_b$ and $u_d$ are the same points (if $u_c < u_d$)
as shown in Figure~\ref{ftp}($b$), otherwise $u_b$ and $u_c$ are the same points (if $u_d < u_c$). In that case either we have a $K_{2,2}$ (forbidden by the {\it path restricted property})
or $u_a$ has a back edge incident to a vertex in $<v_i,v_j>$ violating the {\it path restricted property}. It contradicts to the assumption that $P_1$ and $P_2$ meet at $v_i$.
Therefore, two forward paths originating from a vertex in the same direction never meet each other.
\qed
\begin{corollary} \label{rem2}
From any vertex (not in the forward path $P$), only one edge can be incident to the vertices in the forward path $P$.
\end{corollary}
The stated corollary can be proved by Lemma~\ref{rem1}. Let in a $PRBG$ $G = (U,V,E)$, there exists a forward
path $P$ with the range $\{<u_a,u_b>,<v_c,v_d>\}$ s.t. $u_a,u_b \in U$ and $v_c,v_d \in V$. Let $u_e \in U$ has two edges incident to the vertices in this forward path (assume $v_c$ and $v_d$ w.l.o.g.).
It implies that two forward paths originating from a point in the same direction meet again contradicting to Lemma~\ref{rem1}.
If $u_e \notin <u_a,u_b>$ as shown in Figure~\ref{fcor}(a), then two forward path originating from $u_e$ meet at $u_a$. If $u_e \in <u_a,u_b>$ as shown in Figure~\ref{fcor}(b), then two forward paths originating from $v_d$ meet at $v_c$.
Therefore, from any vertex, only one edge can be incident to the vertices in a forward path.

Let us consider all the forward paths originating from a vertex. These paths could be classified into two sets. The first set consists of all the forward paths visiting to the lower ordered vertices (rightwards) and 
the second set consists of all the forward paths visiting to the higher ordered vertices (leftwards).
Let us consider first the set of the paths visiting rightwards. 
From the subsequent vertices on these paths, multiple paths can originate visiting to the vertices rightwards. These paths never meet with each other (refer to Lemma~\ref{rem1}).
Thus, these forward paths originating from a vertex form a tree. Let $T_r(u)$ denotes such a tree originating from $u$.
Similarly, $T_l(u)$ denotes a tree that consists of all the forward paths originating from $u$ visiting the higher ordered vertices (leftwards).
\begin{lemma}
For any vertex $v$ in a $PRBG$ $G=(U,V,E)$, the subgraph induced by the vertices of $T_r(v)$ has $n-1$ edges where $n$ is the number of vertices spanned by $T_r(v)$.
\end{lemma}
\proof
We show that for any vertex $v$ (let $v \in V$ w.l.o.g) in a $PRBG$, the subgraph induced by the vertices in $T_r(v)$ does not have any edge but the edges in $T_r(v)$.
On the contrary, let there exists an edge $(u_i,v_i) \in E$  s.t. this edge is not present in $T_r(v)$ and the vertices ($u_i \in U$ and $v_i \in V$) are spanned by $T_r(v)$. Recall that two forward paths emerging from a vertex
in the same direction never meet again (refer to Lemma~\ref{rem1}). Therefore, the edge $(u_i,v_i)$ does not belong to any forward path emerging from $v$. Let $u_j \in U$ be the vertex with the highest order incident to $v$.
Note that $u_i$ and $u_j$ are not the same vertices and $u_i < u_j$ (refer to Figure~\ref{u3}(a)). $u_i$ cannot have an edge incident to $v$, otherwise the edge $(u_i,v_i)$ belongs to a forward path originating from $v$ as shown in Figure~\ref{u3}(b).
But there exists a forward path passing through $v$ and $u_i$. Let $v_{i'} \in V$ be the vertex preceding $u_i$ in the forward path from $v$ to $u_i$. Observe
that $v_{i'} < v_i$. Thus, there exists a forward path with the range $\{<u_i,u_j>,<v_{i'},v>\}$. Therefore, the back edge $(u_i,v_i)$ is forbidden by the definition of $PRBGs$.
Thus, it leads to a contradiction to the assumption that there exists an edge between $u_i$ and $v_i$.
\qed
\begin{figure}[ht]
\begin{minipage}[b]{0.5\linewidth}
\begin{center}
\includegraphics[scale=0.35]{udg14}
\caption{Edge $(u_i,v_i)$ is forbidden}
\label{u3}
\end{center}
\end{minipage}
\hspace{0.5cm}
\begin{minipage}[b]{0.4\linewidth}
\centering
\includegraphics[scale=0.6]{udg3}
\caption{Edges in $T_l(v)$}
\label{u4}
\end{minipage}
\end{figure}
\begin{lemma}
 For any vertex $v$ in a $PRBG$ $G = (U,V,E)$, all the forward paths in $T_l(v)$ have disjoint ranges.
\end{lemma}
\proof
Let us assume w.l.o.g. that $v \in V$. Consider two forward paths in $T_l(v)$ originating from $v$. Consider a path $P_1 = (v, u_1, v_1, \ldots)$ as shown in Figure~\ref{u4}.
Also consider the path $P_2 = (v,u_2,v_2, \ldots)$ where $v_1 < v_2$ (for $v_1,v_2 \in V$). Observe that there is a restriction that $u_1 > u_2$ ($u_1,u_2 \in U$), otherwise
the edge $(u_1,v_1)$ is forbidden by the {\it path restricted property}. Similarly, let $u_i \in U$ and $v_i \in V$ be the successive vertices in $P_1$
and let $u_j \in U$ and $v_j \in V$ be the successive vertices in $P_2$. By the {\it path restricted property}, it can be observed that if $v_i < v_j$, then $u_j < u_i$.
Therefore, the ranges of the paths $P_1$ and $P_2$ are disjoint.
\qed
\section{Edge complexity of path restricted ordered bipartite graphs}\label{sc4}
In this section, we study $PRBGs$ for their edge complexity. We also study the edge complexity of these graphs for a special case
when the length of the longest forward path is bounded.
\begin{lemma}[Crossing lemma]\label{pl}
 Consider a $PRBG$ $G = (U, V, E)$ with a separator line $\ell$ partitioning $U$ (resp. $V$) into disjoint subsets $U_1$ and $U_2$ (resp. $V_1$ and $V_2$) 
 s.t. all the vertices in $U_1$ and $V_1$ are placed to the left of $\ell$ and all the vertices in $U_2$ and $V_2$ are placed to the right of $\ell$.
\begin{enumerate}
 \item If every vertex in $U_1$ has an edge incident to it with the other endpoint in $V_1$, then the number of edges between $U_1$ and $V_2$ (crossing $\ell$) is at most $|U_1| + |V_2|$.
 \item If every vertex in $V_1$ has an edge incident to it with the other endpoint in $U_1$, then the number of edges between $V_1$ and $U_2$ (crossing $\ell$) is at most $|V_1| + |U_2|$.
\end{enumerate}
\end{lemma}
\begin{figure}[h]
\begin{center}
 \includegraphics [scale=0.4]{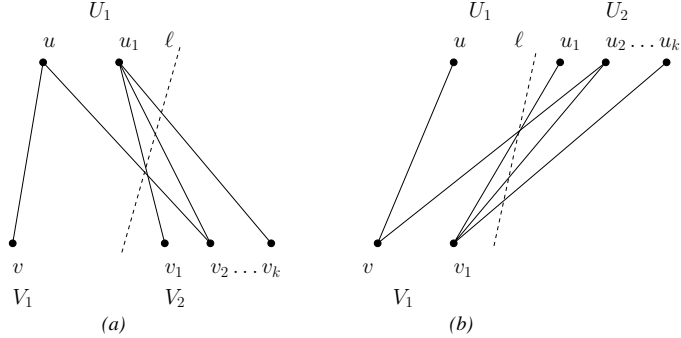}
 \caption{Edges across a partition line}
 \label{u9}
\end{center}
\end{figure}
\proof
An edge crossing the partition line $\ell$ is called the crossing edge.
Let us consider only the vertices (in either of $U_1, U_2, V_1$ and $V_2$) that have more than one crossing edges incident to them.
We give unit charge to all the vertices initially. A vertex can consume its charge to count for an edge.
We show that if every vertex is charged for the leftmost crossing edge incident to it, then all the edges are counted. 

Consider the rightmost vertex $u_1 \in U_1$ (the vertex with the least order in $U_1$) that has crossing edges incident to the vertices $v_1,v_2, \ldots, v_k$ as shown in Figure~\ref{u9}($a$).
We show that any of these vertices except $v_1$ cannot have an edge incident to a vertex in $U_1$ placed to the left of $u_1$. Let us assume on the contrary that $v_2$ has such an edge incident to the vertex $u$. 
By assumption $u$ has an edge incident to a vertex in $V_1$ (say $v \in V_1$), the edge does not intersect $\ell$ and it is placed to the left of it. Since, $v_1$ is placed to the right of $\ell$,
there exists a forward path with the range $\{<u,u_1>,<v,v_2>\}$ and the back edge $(u_1,v_1)$ is forbidden by the {\it path restricted property} since $v_1 \in <v,v_2>$. Thus, it contradicts to the assumption that $v_2$ has an edge incident to $u$.
Since $u_1$ is the rightmost vertex in $U_1$, the vertices $v_2, \ldots, v_k$ have only one crossing edge incident to them. These vertices consume their charges to count the corresponding edges. $u_1$ consumes its charge for the edge $(u_1,v_1)$.
Note that all the crossing edges incident to $u_1$ and its adjacent vertices across $\ell$ (except $v_1$) are counted. Also note that the charge of $v_1$ is still not consumed.
Now. this charging scheme can be applied to the next vertex to the left of $u_1$. Subsequently, this procedure can be applied to all the vertices in $U_1$ from right to left and all the edges are counted.
Thus, if each vertex in $U_1$ and $V_2$ consumes its charge to count the leftmost edge incident to it, all the edges between $U_1$ and $V_2$ are counted.

\begin{figure}[h]
\begin{center} \label{u5}
 \includegraphics [scale=0.4]{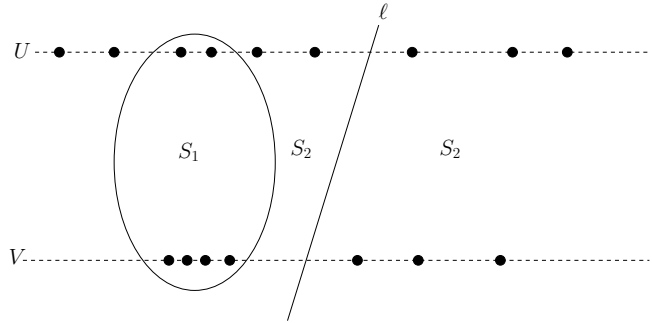}
 \caption{Partition of the point set}
\end{center}
\end{figure}

Similarly for the proof of (2), if a vertex $v_1 \in V_1$ that has crossing edges incident to the vertices $u_1,u_2, \ldots u_k$ as shown in Figure~\ref{u9}($b$), then the vertices $u_2, \ldots, u_k$ cannot have an incident to a vertex
in $V_1$ placed to the left of $v_1$. 
A similar argument can be made to  show that if each vertex in $V_1$ and $U_2$ consumes its charge to count the leftmost edge incident to it, then all the edges between $V_1$ and $U_2$ are counted.
\qed
\begin{theorem} \label{T2}
Any path restricted ordered bipartite graph $G = (U,V,E)$ has at most $n \log n + O(n)$ edges where $n = |U| + |V|$. The bound is tight as there exists a path restricted ordered bipartite graph on $n$ vertices with $\Omega(n \log n)$ edges.
\end{theorem}
\proof
We propose a simple divide and conquer technique to get the desired bound. A partition line $\ell$ is drawn dividing the vertices into two halves. Now, we divide the vertices into two disjoint subsets $S_1$
and $S_2$ as shown in Figure~\ref{u5}. All the vertices in $S_1$ are placed to the left of $\ell$ whereas the vertices in $S_2$ can be placed to both sides of $\ell$.
A simple procedure is used to obtain the partition.
In the partition $V$, the vertices are scanned from left to right. These vertices along with all their neighbors in $U$ are included in $S_1$. The process is stopped when $S_1$ has at least $\frac{n}{2}$ vertices.
Consider the situation when before scanning a vertex $v_i$, there are less than $\frac{n}{2}$ vertices in $S_1$. After $v_i$ is scanned, there are more than $\frac{n}{2}$ vertices in $S_1$. Note that all the new vertices added
to $S_1$ while scanning $v_i$ are the pendant vertices within $S_1$, i.e. these vertices have only one edge incident to them in the subgraph induced on the vertices in $S_1$. 
All other edges incident to these vertices cross $\ell$. These vertices are called the {\it terminal vertices}. The partition obtained by this procedure has the following properties.
\begin{enumerate}
 \item If any edge incident to a vertex in $S_1$ has its other end point to the left
of $\ell$, then the corresponding vertex must be in $S_1$.
 \item For any vertex in $S_1$, there is at least one edge incident to another vertex in $S_1$, i.e. both the vertices defining the edge are placed to the left of $\ell$.
\end{enumerate}

Let us now consider the edges with one end point in $S_1$ and the other end point in $S_2$. All such edges must cross the line $\ell$ by property (1) of the partition.
Lemma~\ref{pl} can be applied to count such edges due to property (2) of the partition.
By Lemma~\ref{pl}, the maximum number of these edges is at most the summation of the number of vertices in $S_1$ and the number of vertices (in $S_2$) that are placed to the right of $\ell$.
Thus, the number of such edges is at most $n-1$. Let $\mathcal{T}(P)$ denote the maximum number of edges a $PRBG$ on a vertex set $P$ can have, then $\mathcal{T}(S_1 \cup S_2) \le \mathcal{T}(S_1) + \mathcal{T}(S_2) + n-1$.
Terminal vertices can be dropped from $S_1$ as they have only one edge incident to them. Thus, both the partitions $S_1$ and $S_2$ have at most $\frac{n}{2}$ vertices.
Now the same procedure can be independently applied to count the edges in $S_1$ and $S_2$ recursively. Thus, $\mathcal{T}(U \cup V) = n \log n + O(n)$. It proves that the number of edges in $G$ is at most $n \log n + O(n)$.

A matching lower bound can be obtained by a similar 0-1 matrix proposed in ~\cite{tds}. We present the matrix $\mathcal{A}$ (shown in Figure~\ref{mata}) and construct a bipartite graph using it.
Rows of the matrix represent the adjacencies of the vertices $u_i \in U$. The columns represent adjacencies for the vertices $v_i \in V$. Let the matrix
have $n$ number of rows and columns. The entry corresponding
to the row $i$ and column $j$, $\mathcal{A}(i,j)$ is 1 if $i+j-n = 2^k$ for some integer $k$. 
Now, we show that the bipartite graph corresponding to this adjacency matrix is a $PRBG$. The entries in the adjacency matrix corresponding to a forward path form
a stair case pattern as shown in the matrix depicted in Figure~\ref{mats}.
\begin{figure}[ht]
\begin{minipage}[b]{0.5\linewidth}
\begin{center}
$\begin{matrix}
\begin{bmatrix}
\; & \; & \; & \; & \; & \; & \; & 1\\
\; & \; & \; & \; & \; & \; & 1 & 1\\
\; & \; & \; & \; & \; & 1 & 1 & 0\\
\; & \; & \; & \; & 1 & 1 & 0 & 1\\
\; & \; & \; & 1 & 1 & 0 & 1 & 0\\
\; & \; & 1 & 1 & 0 & 1 & 0 & 0\\
\; & 1 & 1 & 0 & 1 & 0 & 0 & 0\\
1 & 1 & 0 & 1 & 0 & 0 & 0 & 1\\
\end{bmatrix}
\end{matrix}$
\caption{The matrix $\mathcal{A}$}
\label{mata}
\end{center}
\end{minipage}
\hspace{0.5cm}
\begin{minipage}[b]{0.4\linewidth}
\centering
$\begin{matrix}
\begin{bmatrix}
1 & 1\\
\; & 1 & 1\\
\; & \; & 1 & 1\\
\; & \; & \; & 1 & 1\\
\; & 1  & \; & \; & 1 & 1\\
\; & \; & \; & \; & \; & 1\\
\end{bmatrix}
\end{matrix}$\caption{The matrix $\mathcal{S}$}
\label{mats}
\end{minipage}
\end{figure}

Let us consider a matrix $\mathcal{S}$ with such a stair case pattern from its top left corner to the bottom right corner and not all other entries in $\mathcal{S}$ are zeros (refer to Figure~\ref{mats}).
While the stair case pattern corresponds to a forward path, observe that any other 1 entry in $\mathcal{S}$ corresponds to a back edge for this forward path.
Therefore, $\mathcal{S}$ does not represent the adjacency matrix of a $PRBG$ by the {\it path restricted property}. We show that $\mathcal{S}$ is forbidden as a sub matrix in $\mathcal{A}$.
Let us assign an index to each 1 entry in $\mathcal{A}$. Recall that $\mathcal{A}(i,j)$ is 1 if $i+j-n = 2^k$ for some integer $k$. $k$ is the index
assigned to the corresponding 1 entry. Let $i_1$ and $i_2$ be two rows s.t. $\mathcal{A}(i_1,j) = \mathcal{A}(i_2,j) = 1$ where $i_1 < i_2$. Let $k_1$ and $k_2$ be
the indices for the entries $\mathcal{A}(i_1,j)$ and $\mathcal{A}(i_2,j)$ respectively. Observe that $k_2 > k_1$. Let the previous 1 entry in row  $i_2$ (with index $k_2-1$)
has column index $j_2$. Observe that in row $i_1$, the column index for any 1 entry is greater than $j_2$. Consider a row $i_0 < i_1$ s.t. $\mathcal{A}(i_0,j') = \mathcal{A}(i_1,j') = 1$.
If the preceding 1 entry to $ \mathcal{A}(i_1,j')$ in row $i_1$ is at column $j_1$, then the column index for any 1 entry in row $i_0$ is greater than $j_1$. Consider the vertices in 
a froward path and the submatrix induced by these vertices. The above argument implies that in this submatrix there does not exist a 1 entry to the left of the stair case pattern
corresponding to the forward path. A similar argument shows that there does not exist a 1 entry to the right of the stair case pattern corresponding to the forward path. Thus, the sub matrix
$\mathcal{S}$ is forbidden in $\mathcal{A}$. Thus, $\mathcal{A}$ represents a $PRBG$. Observe that $\mathcal{A}$ has $\Omega(n \log n)$ number of 1 entries.
\qed
\begin{theorem}
 Any unit distance graph on a convex point set with $n$ points has at most $2n \log n + O(n)$ edges.
\end{theorem}
\proof
Recall from Section~\ref{sc2} that a $UDG$ on a convex point set can be decomposed into two graphs in $G_{UDG}$ by removing at most $3n$ edges.
A graph in $G_{UDG}$ is also a $PRBG$ that has at most $n \log n + O(n)$ edges. Thus, it concludes that any unit distance graph on a convex point set
has at most $2n \log n + O(n)$ edges.
\qed

\begin{figure}
\begin{center} 
 \includegraphics [scale=0.4]{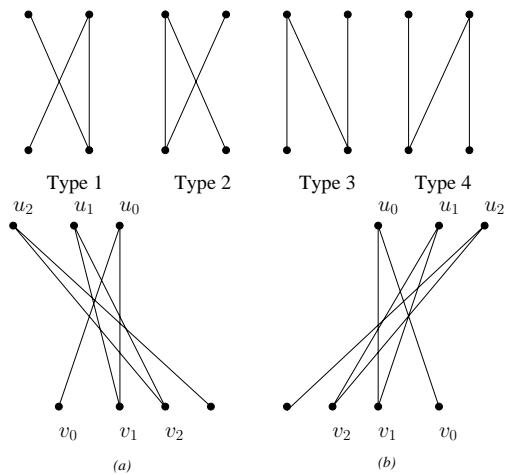}
 \caption {Forbidden patterns and resulting graphs}
\label{put}
\end{center}
\end{figure}

Similarly, the following theorem can be established for the $LGGs$.
\begin{theorem}
 Any locally Gabriel graph on a convex point set with $n$ points has at most $2n \log n + O(n)$ edges.
\end{theorem}
Now, we focus for the edge complexity of $PRBGs$ that satisfy some specific properties on the paths. 
\begin{theorem} \label{p3}
 There can be four kind of paths with length 3 (denoted by $P_3$) as shown in Figure~\ref{put}. If any of the $P_3$ is forbidden in a path restricted ordered bipartite graph $G=(U,V,E)$,
 then $G$ is acyclic.
\end{theorem}
\proof
Let us consider all these four cases one by one. Consider the $P_3$ of Type 1. Consider the left most vertex in $U$ (with the highest order) and the tree $T_r(u)$.
If Type 1 $P_3$ is forbidden, then it implies that the vertices spanned by this tree have no more adjacencies. Similarly, rest of the vertices are part of some tree and the
graph is acyclic. Now let us consider the situation where Type 2 $P_3$ is forbidden. Consider the rightmost vertex in $U$ (with the least order) and all the vertices in $T_l(u)$.
If Type 2 $P_3$ is forbidden, then it implies that the vertices spanned by this tree have no more adjacencies. Similarly, rest of the vertices are part of some tree and the
graph is acyclic.

Now let us consider the case when Type 3 is forbidden. 
Let $v_0$ be the right most vertex in $V$ (with the highest order) and an edge $(u_0,v_0)$ incident to it for some $u_0 \in U$. Let $v_1$ be a vertex with an edge incident to $u_0$, note that $v_1 < v_0$.
Consider any edge incident to $v_1$ (say $(v_1,u_1)$). Since Type 3 is forbidden, $u_1 > u_0$. Furthermore, an edge incident to $u_1$ (say $u_1,v_2)$) implies that $v_2$ has lower order than $v_0$ and $v_1$.
Similarly, for any vertex $u_2$ with an edge incident to $v_2$ implies that $u_2$ has higher order than $u_0$ and $u_1$. Thus, for any path emerging from $v_0$ order of the vertices increases monotonically
in $U$ and decreases monotonically in $V$ as shown in Figure~\ref{put}($a$), Thus, it can be concluded that the graph is acyclic.
A symmetric argument shows that the graph is acyclic when Type 4 $P_3$ is forbidden. Let $v_0$ be the vertex with least order here. On any path emerging from this vertex, the order
of vertices monotonically decreases in $U$ and monotonically increases in $V$ as shown in Figure~\ref{put}($b$).
Thus, it proves that a $PRBG$ is acyclic if any of the four kinds of $P_3$ is forbidden.
\qed

Now, we show that an improved bound on the number of edges when the length of the longest forward path is bounded.

\begin{figure}
\begin{center}
 \includegraphics [scale=0.3]{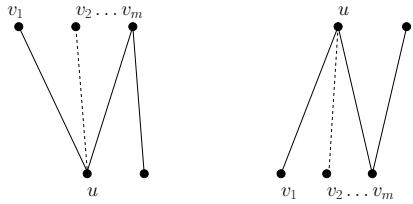}
 \caption{Longest path of length 3 in a graph}
 \label{u7}
\end{center}
\end{figure}

\begin{lemma}
 If length of the longest forward path in a path restricted ordered bipartite graph $G = (U,V,E)$ is at most by $k$, then the graph has
$O(k(|U|+|V|))$ edges.
\end{lemma}
\proof
First, we show that the claim holds for $k = 3$. Two possible kind of forward paths of length 3 are shown in Figure~\ref{u7}. Let $u$ has the edges incident to $v_1, v_2, \ldots, v_m$. Except $v_m$, none of these
vertices can have an edge incident to the left of $u$ otherwise $k > 3$. Therefore, if the left most incident edge is deleted for every vertex, length of the longest forward path reduces to two.
Then, the resulting graph has at most $|U| + |V| - 1$ edges (refer to Theorem~\ref{p3}). Thus, it implies that $G$ has at most $2(|U| + |V|)-1$ edges. Now, let us consider the generic case for any value of $k$.
Iterating the same procedure (deleting the left most edge for every vertex) reduces the length of the longest forward path by one. After $k-2$ iterations of deleting the left most edge for each vertex, the resultant graph
does not have a forward path of length three. Thus, the original graph has $O(k(|U|+|V|))$ edges.
\qed
\section{Hierarchy of various graph classes}\label{sc5}
In this section, we study the relationship amongst various graph classes. First we show that Class $G_{UDG}$ is a strict sub class of the class $G_{LGG}$. Then, 
we show that class $G_{LGG}$ is a strict sub class of the generic path restricted ordered bipartite graphs. We also show that the class of $UDGs$ on convex point sets
is a strict sub class of the $LGGs$ on convex point sets.
\begin{lemma} \label{ul}
 Class $G_{UDG}$ is a strict sub class of the class $G_{LGG}$.
\end{lemma}
\begin{figure}[ht]
\begin{minipage}[b]{0.5\linewidth}
\centering
\includegraphics[scale=0.3]{udgf}
\caption{A forbidden $G_{UDG}$}
\label{u6}
\end{minipage}
\hspace{0.5cm}
\begin{minipage}[b]{0.4\linewidth}
\centering
\includegraphics[scale=0.35]{udg15}
\caption{A forbidden $G_{LGG}$}
\label{u8}
\end{minipage}
\end{figure}
\proof
We show a simple example of a graph that is forbidden in the class $G_{UDG}$ and can be easily embedded as an $G_{LGG}$. Consider the graph
shown in Figure~\ref{u6}, we show that this graph cannot be embedded as  $G_{UDG}$.
In the quadrilateral $u_1v_3v_5u_3$, by the definition of $G_{UDG}$ and convexity, $\angle v_3u_1u_2 < \frac{\pi}{2}$. It can be observed by convexity that $\angle v_3u_1u_3 < \frac{\pi}{2}$.
By the property of isosceles triangles, $\angle u_1v_3v_5$ and $\angle v_5u_3u_1$ are acute. Therefore, in the quadrilateral $u_1v_3v_5u_3$, $\angle u_3v_5v_3$ is greater than $\frac{\pi}{2}$.
By convexity, $\angle u_3v_5v_4$ is greater than $\frac{\pi}{2}$. Therefore, $\overline{v_4u_3} > \overline{v_5u_3}$. Since $ \overline{v_5u_3}$ has unit length, $\overline{v_4u_3}$ has length more than unity.
The locus of the points equidistant from $u_2$ and $u_3$ is the perpendicular bisector to the line joining these points as shown in Figure~\ref{u6}.
Observe that $\overline{v_1u_3} > \overline{v_1u_2}$ and the length of $\overline{v_1u_3}$ is greater than unity. Also observe that $\overline{u_3v_3}$ is greater than unity since $\angle u_3v_5v_3 \ge \frac{\pi}{2}$.
Since the distance of $u_3$ from both the vertices $v_1$ and $v_3$ is greater than unity, no vertex $v_2$ can be chosen (between $v_1$ and $v_3$) with unit distance from $u_3$ s.t. all the points are in convex
position. Observe that this graph can be easily embedded as an $G_{LGG}$.
\qed
\begin{figure}
\begin{center} 
 \includegraphics [scale=0.3]{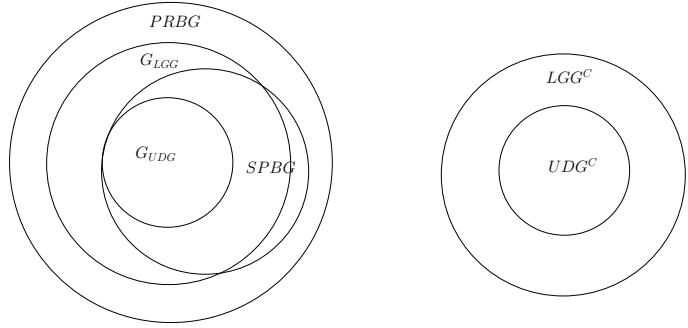}
 \caption{Hierarchy of various graphs}
 \label{u10}
\end{center}
\end{figure}
\begin{definition}
A $PRBG$ $G = (U, V, E)$ is called strictly path restricted ordered bipartite graph ($SPBG$), if two vertices $v_1,v_2 \in V$ s.t. $v_1 < v_2$ are spanned by some tree $T_r(v), v \in V$ 
and $u_1$ and $u_2$ be the vertices preceding $v_1$ and $v_2$ respectively in the forward paths from $v$ to $v_1$ and $v_2$ and $u_1 < u_2$, then $u_1$ and $u_2$ cannot have edges incident
to the vertices $v'_1$ and $v'_2$ (not spanned by $T_r$(v)) s.t. $v'_1 < v'_2$.
\end{definition}
\begin{remark}
 In a strictly path restricted ordered bipartite graph $G = (U,V,E)$ if two vertices $u_1 \in U$ and $v_1 \in V$ are spanned by some tree $T_l(u)$, then there does not exist an edge between $u_1$ and $v_1$. 
\end{remark}
It can be observed that a $UDG$ on a convex point set can be represented as strictly path restricted ordered bipartite graph (refer to Lemma~\ref{ul}).

\begin{lemma}
Class $G_{LGG}$ is a strict sub class of the generic path restricted ordered bipartite graphs. 
\end{lemma}
\proof
We show a simple example of a graph that is a $PRBG$ and forbidden in the class $G_{LGG}$. 
Consider the graph shown in Figure~\ref{u8} . The graph does not violate the {\it path restricted property} of the $PRBGs$. It can be argued that the graph cannot be represented as $G_{LGG}$. Recall that in an $LGG$ if 
there exist edges $(u,v_1)$ and $(u,v_2)$, then $\angle uv_1v_2 < \frac{\pi}{2}$ and $\angle uv_2v_1 < \frac{\pi}{2}$. Therefore, all the four angles
$\angle u_1v_1v_2, \angle v_1u_1u_2, \angle v_4u_4u_3$ and $\angle u_4v_4v_3$ need to be acute in an $LGG$. By convexity, $\angle v_1u_1u_4, \angle u_1u_4v_4, \angle u_4v_4v_1$
and $\angle u_4v_1u_1$ are acute. That is not possible because at least one angle in the quadrilateral $u_1u_4v_4v_1$ must be obtuse. Thus, this graph cannot be represented
as $G_{LGG}$ or a locally Gabriel graph on a convex point set.
\qed
Therefore, a strict hierarchy can be established among three families of the graphs. $G_{UDG}$ is a strict sub class of the class of the graphs represented by $G_{LGG}$. Furthermore,
$G_{LGG}$ is a strict subclass of the ordered bipartite graphs that satisfy path restricted property. The family of strictly path restricted ordered bipartite graphs ($SPBG$) is an obvious sub class of the 
generic $PRBGs$. The hierarchy is shown pictorially in Figure~\ref{u10}. Though a $G_{UDG}$ can be represented as a $SPBG$,
it is not known whether there is an equivalence between these two classes of graphs. There exist $G_{LGG}$ not belonging to the class of $SPBGs$.
It is not clear whether all $SPBGs$ can be represented as $G_{LGG}$.

Let $UDG^C$ and $LGG^C$ be the classes of all the unit distance graphs and the locally Gabriel graphs on convex point sets. It can be observed in Figure~\ref{u6}, if the points 
$v_3$ and $v_4$ coincide then this graph cannot be embedded as unit distance graphs on a convex point but can be embedded as a locally Gabriel graph on a convex point set. It also
establishes that the class $UDG^C$ is a strict subclass of $LGG^C$.
\section{Concluding Remarks}
In this note, we defined a family of bipartite graphs known as the path restricted ordered bipartite graphs. We also showed that these graphs can be obtained from various geometric graphs
on convex point sets. We studied various structural properties of these graphs and showed that a path restricted ordered bipartite graph on $n$ vertices has $O(n \log n)$ edges and this
bound it tight. The same upper bound is already known for the unit distance graphs and the locally Gabriel graphs on convex point sets. However, the best known lower bound known
to the edge complexity on these graphs for convex point sets is $\Omega(n)$. The major challenge
in this direction of work is to bridge this gap. It is interesting to know whether a $o(n \log n)$ upper bound can be established for the strictly path restricted ordered
bipartite graphs as it also improves the upper bound for the unit distance graphs on convex point sets. The problem of bridging the gap in the bounds also remains open for the
locally Gabriel graphs on a convex point set.\\
\paragraph{Acknowledgement:} The authors are thankful to Subramanya Bharadwaj for useful comments towards the proof of Theorem~\ref{T2}.
\bibliographystyle{amsplain}
\bibliography{references}

\providecommand{\bysame}{\leavevmode\hbox to3em{\hrulefill}\thinspace}
\providecommand{\MR}{\relax\ifhmode\unskip\space\fi MR }
\providecommand{\MRhref}[2]{%
  \href{http://www.ams.org/mathscinet-getitem?mr=#1}{#2}
}
\providecommand{\href}[2]{#2}
\begin{thebibliography}{10}

\bibitem{csc}
Bernardo~M. {\'A}brego and Silvia Fern{\'a}ndez-Merchant, \emph{The unit
  distance problem for centrally symmetric convex polygons}, Discrete {\&}
  Computational Geometry \textbf{28} (2002), no.~4, 467--473.

\bibitem{beck}
J{\'o}zsef Beck and Joel Spencer, \emph{Unit distances}, J. Comb. Theory, Ser.
  A \textbf{37} (1984), no.~3, 231--238.

\bibitem{sas}
Peter Brass, Gyula Károlyi, and Pavel Valtr, \emph{A {T}ur{\'a}n-type extremal
  theory of convex geometric graphs}, Discrete and Computational Geometry,
  Algorithms and Combinatorics, vol.~25, Springer Berlin Heidelberg, 2003,
  pp.~275--300 (English).

\bibitem{bp}
Peter Bra{\ss} and J{\'a}nos Pach, \emph{The maximum number of times the same
  distance can occur among the vertices of a convex n-gon is {O}(n log n)}, J.
  Comb. Theory, Ser. A \textbf{94} (2001), no.~1, 178--179.

\bibitem{Capoy}
Vasilis Capoyleas and J{\'a}nos Pach, \emph{A {T}ur{\'a}n-type theorem on
  chords of a convex polygon}, Journal of Combinatorial Theory, Series B
  \textbf{56} (1992), no.~1, 9--15.

\bibitem{han}
Herbert Edelsbrunner and Péter Hajnal, \emph{A lower bound on the number of
  unit distances between the vertices of a convex polygon.}, J. Comb. Theory,
  Ser. A \textbf{56} (1991), no.~2, 312--316.

\bibitem{Erd86}
P.~Erd\H{o}s, \emph{On some metric and combinatorial geometric problems},
  Discrete Mathematics \textbf{60} (1986), no.~0, 147--153.

\bibitem{Erdudg}
Paul Erd\H{o}s, \emph{On sets of distances of n points}, The American
  Mathematical Monthly \textbf{53} (1946), no.~5, pp. 248--250.

\bibitem{Fb92}
Peter~C. Fishburn and James~A. Reeds, \emph{Unit distances between vertices of
  a convex polygon}, Comput. Geom. \textbf{2} (1992), 81--91.

\bibitem{fox}
Jacob Fox, J{\'a}nos Pach, and CsabaD T{\'o}th, \emph{Tur{\'a}n-type results
  for partial orders and intersection graphs of convex sets}, Israel Journal of
  Mathematics \textbf{178} (2010), no.~1, 29--50 (English).

\bibitem{furedi}
Zolt{\'a}n F{\"u}redi, \emph{The maximum number of unit distances in a convex
  {\it n}-gon}, J. Comb. Theory, Ser. A \textbf{55} (1990), no.~2, 316--320.

\bibitem{H92}
Zolt{\'a}n F{\"u}redi and P{\'e}ter Hajnal, \emph{Davenport-schinzel theory of
  matrices}, Discrete Mathematics \textbf{103} (1992), no.~3, 233--251.

\bibitem{ggg}
Ruben~K. Gabriel and Robert~R. Sokal, \emph{A new statistical approach to
  geographic variation analysis}, Systematic Zoology \textbf{18} (1969), no.~3,
  259--278.

\bibitem{szem}
S.~J{\'o}zsa and E.~Szemer{\'e}di, \emph{The number of unit distances on the
  plane}, Infinite and finite sets, Coll. Math. Soc. J. Bolyai \textbf{10}
  (1973), 939--950.

\bibitem{yyy}
Sanjiv Kapoor and Xiang-Yang Li, \emph{Proximity structures for geometric
  graphs}, International Journal of Computational Geometry and Applications,
  vol.~20, 2010, pp.~415--429.

\bibitem{Keszegh09}
Bal{\'a}zs Keszegh, \emph{On linear forbidden submatrices}, J. Comb. Theory,
  Ser. A \textbf{116} (2009), no.~1, 232--241.

\bibitem{perles}
Y.S. Kupitz and M.A. Perles, \emph{Extremal theory for convex matchings in
  convex geometric graphs}, Discrete {\&} Computational Geometry \textbf{15}
  (1996), no.~2, 195--220 (English).

\bibitem{ram}
Gy. Károlyi, J.~Pach, G.~Tóth, and P.~Valtr, \emph{Ramsey-type results for
  geometric graphs, ii}, Discrete {\&} Computational Geometry \textbf{20}
  (1998), no.~3, 375--388 (English).

\bibitem{crss}
J.~Pach, F.~Shahrokhi, and M.~Szegedy, \emph{Applications of the crossing
  number}, Algorithmica \textbf{16} (1996), no.~1, 111--117 (English).

\bibitem{pach}
J\'{a}nos Pach and G\'{a}bor Tardos, \emph{Forbidden patterns and unit
  distances}, Proceedings of the twenty-first annual symposium on Computational
  geometry (New York, NY, USA), SCG '05, ACM, 2005, pp.~1--9.

\bibitem{pet10}
Seth Pettie, \emph{On nonlinear forbidden 0-1 matrices: A refutation of a
  f{\"u}redi-hajnal conjecture}, SODA, 2010, pp.~875--885.

\bibitem{ps04}
Rom Pinchasi and Shakhar Smorodinsky, \emph{On locally {D}elaunay geometric
  graphs}, Proceedings of the twentieth annual symposium on Computational
  geometry (New York, NY, USA), ACM, 2004, pp.~378--382.

\bibitem{bb}
Joel Spencer, Endre Szemer{\'e}di, and William~T. Trotter, \emph{Unit distances
  in the euclidean plane}, pp.~293--308, Academic Press, 1984.

\bibitem{suk}
Andrew Suk, \emph{Turan and ramsey type problems on geometric objects}, Ph.D.
  thesis, New York, NY, USA, 2011, AAI3445328.

\bibitem{Szk}
L{\'a}szl{\'o}~A. Sz{\'e}kely, \emph{Crossing numbers and hard {E}rd{\H o}s
  problems in discrete geometry}, Comb. Probab. Comput. \textbf{6} (1997),
  no.~3, 353--358.

\bibitem{tds}
G{\'a}bor Tardos, \emph{On 0-1 matrices and small excluded submatrices}, J.
  Comb. Theory, Ser. A \textbf{111} (2005), no.~2, 266--288.

\bibitem{tth}
G.~T{\'o}th and P.~Valtr, \emph{Geometric graphs with few disjoint edges},
  Discrete {\&} Computational Geometry \textbf{22} (1999), no.~4, 633--642
  (English).

\bibitem{toth}
G{\'e}za T{\'o}th, \emph{Note on geometric graphs}, Journal of Combinatorial
  Theory, Series A \textbf{89} (2000), no.~1, 126--132.

\bibitem{turan}
Paul Tur{\'a}n, \emph{On an extremal problem in graph theory}, Matematikai
  {\'e}s Fizikai Lapok \textbf{48} (1941), 436--452.

\bibitem{pv}
Pavel Valtr, \emph{Graph drawing with no k pairwise crossing edges}, Graph
  Drawing (Giuseppe DiBattista, ed.), Lecture Notes in Computer Science, vol.
  1353, Springer Berlin Heidelberg, 1997, pp.~205--218.

\end{thebibliography}
\end{document}